\UseRawInputEncoding

\documentclass[
reprint,
superscriptaddress,
amsmath,
amssymb,
aps,
]{revtex4-2}

\usepackage{graphicx}
\usepackage{dcolumn}
\usepackage{bm}
\usepackage{physics}
\usepackage{siunitx}
\usepackage{romannum}
\usepackage{amssymb}
\usepackage{hyperref}
\hypersetup{colorlinks= true,linkcolor=blue}
\begin{document}

\title{Microwave spectroscopy of few-carrier states in bilayer graphene quantum dots}

\author{Max J. Ruckriegel}
\author{Christoph Adam}
\author{Rebecca Bolt}
\affiliation{Laboratory for Solid State Physics, ETH Z\"urich, CH-8093 Z\"urich, Switzerland}
\author{Chuyao Tong}
\author{David Kealhofer}
\author{Artem O. Denisov}
\affiliation{Laboratory for Solid State Physics, ETH Z\"urich, CH-8093 Z\"urich, Switzerland}
\author{Mohsen Bahrami Panah}
\affiliation{Laboratory for Solid State Physics, ETH Z\"urich, CH-8093 Z\"urich, Switzerland}
\affiliation{Quantum Center, ETH Z\"urich, CH-8093 Z\"urich, Switzerland}
\author{Kenji Watanabe}
\affiliation{Research Center for Electronic and Optical Materials, National Institute for Materials Science, 1-1 Namiki, Tsukuba 305-0044, Japan}
\author{Takashi Taniguchi}
\affiliation{Research Center for Materials Nanoarchitectonics, National Institute for Materials Science,  1-1 Namiki, Tsukuba 305-0044, Japan}
\author{Thomas Ihn}
\author{Klaus Ensslin}
\affiliation{Laboratory for Solid State Physics, ETH Z\"urich, CH-8093 Z\"urich, Switzerland}
\affiliation{Quantum Center, ETH Z\"urich, CH-8093 Z\"urich, Switzerland}

\date{\today}

\begin{abstract}
Bilayer graphene is a maturing material platform for gate-defined quantum dots that hosts long-lived spin and valley states.
Implementing solid-state qubits in bilayer graphene requires a fundamental understanding of such confined electronic systems.
In particular, states of two and three carriers, for which the exchange interaction between particles plays a crucial role, are a cornerstone for qubit readout and manipulation.
Here we report on the spectroscopy of few-carrier states in bilayer graphene quantum dots, using circuit quantum electrodynamics (cQED) techniques that offer substantially improved energy resolution compared to standard transport techniques.
Measurements using a superconducting high-impedance resonator capacitively coupled to the double quantum dot reveal dispersive features of two and three electron states, enabling the detection of Pauli spin and valley blockade and the characterization of the spin-orbit gap at zero magnetic field.
The results deepen our understanding of few-carrier spin and valley states in bilayer graphene quantum dots and demonstrate that cQED techniques are a powerful state-selective probe for semiconductor nanostructures.
    
\end{abstract}

\maketitle
%
%
%
\section{Introduction}
Bilayer graphene (BLG) is an attractive material platform to host quantum dots (QDs).
In addition to their intrinsic spin, individual charge carriers in the atomically thin semiconductor BLG feature the so-called valley pseudospin, which represents a reliably tunable and robust quantum degree of freedom (d.o.f.) \cite{kurzmann_excited_2019, tong_tunable_2020, tong_pauli_2022, garreis_long-lived_2023, Denisov2025}.
The prospect of encoding quantum states in both the spin and valley quantum numbers therefore presents a unique opportunity for BLG QDs. 
Experiments have shown that these properties give rise to long qubit lifetimes \cite{garreis_long-lived_2023, Denisov2025}.

To achieve, in BLG-based qubits, the level of control, reproducibility, and scalability of spin qubits based on industrially standard materials like silicon, precise measurement of the multi-carrier spectrum is essential.
While the state spectrum of a single carrier confined in a single BLG QD is well studied \cite{kurzmann_excited_2019, Duprez2024, Denisov2025, Adam2025}, a realistic qubit implementation will likely employ multiple carriers distributed over more than one dot \cite{Burkard2023}.
For example, in a double quantum dot (DQD) system, interactions between carriers can be exploited for state-readout or manipulation. 
Previous experiments typically inferred state properties from bias-dependent transport through a DQD, revealing valley- or spin-polarized ground states via Pauli blockade \cite{tong_pauli_2022, banszerus_particlehole_2023, tong_pauli_2023, Tong2024}.
However, such DQD transport measurements have mostly focused on high magnetic fields \cite{Moeller2021, tong_pauli_2022, Tong2024, Moeller2024}, where state degeneracies are lifted and energy splittings are large as a result of the valley- and spin-Zeeman effects.
Owing to the inherent limitations in energy resolution of transport based on sequential tunneling, state properties typically needed to be extrapolated to low magnetic fields.

In this work, we study low-energy valley and spin states in a BLG DQD at zero magnetic field.
By measuring the DQD using a circuit quantum electrodynamics (cQED) architecture, we achieve a remarkable energy resolution far surpassing that of dc transport measurements.
A capacitively coupled on-chip superconducting resonator is sensitive to electric dipole transitions between DQD states \cite{frey_dipole_2012, Ruckriegel2024}. 
Finite source-drain bias introduces a non-equilibrium population that enhances the visibility of excited state dipole transitions \cite{Mi2017High, Viennot2014}.
With this technique, we observe zero-field energy gaps in the DQD states due to Kane-Mele SO interaction.
These states could so far not be clearly resolved in direct transport measurements of DQDs \cite{Banszerus2021, tong_pauli_2022, banszerus_particlehole_2023, Tong2024, Moeller2024}. 
In contrast to transport, the energy resolution of microwave detection is not limited by the finite electronic temperature of the leads, co-tunneling, or phonon interactions \cite{tong_pauli_2022, Tong2024, Moeller2024}.
In our experiments, the state-dependent coupling due to selection rules between states is evident from microwave detection of Pauli blockade via the resonator signal.
The presented state-selectivity makes this technique attractive for fast readout of singlet-triplet qubits \cite{zheng_rapid_2019, landig_microwave-cavity-detected_2019, Ruckriegel2024}.

\section{Dipole coupling of valley and spin DQD states}
%

\begin{figure}[t!]
\includegraphics{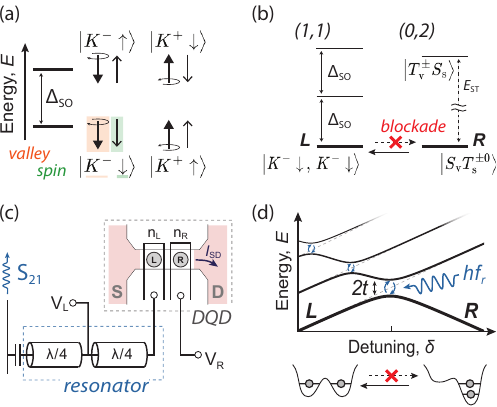}
\caption{\label{fig1} 
(a) Single-carrier states in BLG are labeled by a valley and spin. 
Spin-orbit coupling $\Delta_\mathrm{SO}$ separates the four possible states into two degenerate Kramers pairs.
States with parallel alignment of valley and spin magnetic moments (symbolized by arrows) are lowered in energy, while anti-parallel alignment raises their energy. 
(b) Two-carrier DQD states at the $(1,1) \leftrightarrow (0,2)$ interdot transition.
The valley-polarized $(1,1)$ ground state is incompatible with the valley-singlet $(0,2)$ ground state, leading to Pauli blockade.   
(c) Device schematic of the hybrid device.
A high-impedance resonator is probed through a feedline in a notch-type geometry. 
Its center conductor is connected to the left plunger gate.
The DQD device is fabricated from a van der Waals hetero-structure with BLG encapsulated in hBN.
(d) Energy diagram of the two-carrier DQD states as a function of detuning $\delta$ with finite tunneling coupling $t$ that hybridizes the electronic states.
Their transition dipole moment interacts dispersively with the microwave resonator of energy $h f_\mathrm{r}<2t$.}
\end{figure}

In BLG, intrinsic Kane-Mele SO coupling $\Delta_\mathrm{SO}$ partially lifts the fourfold degeneracy of the single-particle ground state at zero magnetic field.
Its effect is a lowering or raising of the state energies depending on the alignment of valley and spin that results in the degenerate Kramers pairs $\ket{K^- \downarrow}$, $\ket{K^+ \uparrow}$ and $\ket{K^- \uparrow}$, $\ket{K^+ \downarrow}$ illustrated in Figure \ref{fig1} (a).
The zero-field SO energy gap has consistently been characterized to be $\Delta_\mathrm{SO} \approx 60-\SI{80}{\micro\electronvolt}$\cite{kurzmann_excited_2019, Banszerus2021, Duprez2024, Denisov2025}.

SO interaction also affects the spectrum of two and three carriers.
The exact characteristics of two- and three-carrier states in BLG are furthermore known to be influenced by a complex interplay between long- and short-range exchange and their orbital d.o.f. \cite{Knothe2020}.
The resulting valley- and spin-polarized states are observed by Pauli blockade [see Fig.~\ref{fig1} (b)] \cite{tong_pauli_2022, Tong2024}, an effect that provides the valley- and spin-to-charge conversion necessary for operating valley or spin qubits in BLG. 
In this context, a thorough spectroscopic characterization of low-lying excited states is crucial, in particular at low magnetic fields where energy splittings fall in the experimentally desirable range below $\sim \SI{12}{\giga\hertz}$.
For spin qubits in silicon QDs, for example, low-lying valley states present major challenges as they lead to considerable decoherence and complicate initialization, manipulation and readout \cite{Yang2013, Kawakami2014}. 

In order to measure the low-energy spectrum of DQD states beyond the capabilities of direct transport, cQED techniques can be employed \cite{mi_high_2017, landig_microwave-cavity-detected_2019, borjans_probing_2021}.
Capacitive coupling of the DQD to an on-chip microwave resonator [see Figure \ref{fig1} (c)] directly senses the electric dipole moment of interdot transitions that hybridize through a finite tunnel coupling $t$ \cite{frey_dipole_2012, Ruckriegel2024}.
The interaction is dispersive with $2t > h f_\mathrm{r}$, where $f_\mathrm{r}$ is the resonator frequency, such that dipole transitions are virtual and no energy is exchanged.
Furthermore, electric dipole coupling is only sensitive to the non-blocked transitions between valley- or spin-polarized states.

The effect of the DQD states on the resonator is condensed into its electric susceptibility $\chi$, which describes the system's polarizability in response to the electric field of the resonator.
For a transition between the electronic states $i$ and $j$, the susceptibility is given by \cite{Burkard2016, mi_high_2017}
\begin{equation}
    \chi_{ij} = g_{ij} \cdot \frac{\Delta p_{ij}}{2\pi f_\mathrm{r} - E_{ij}/\hbar + i \gamma}.
\end{equation}
The effective coupling strength $g_{ij} = g_0 d_{ij}$ is proportional to the bare coupling strength $g_0$ and the transition dipole moment $d_{ij}(\delta)$.
This detuning-dependent dipole moment is largest when DQD states hybridize as a result of a finite tunnel coupling $t_{ij}$.
Detuning and tunnel coupling also determine the energy difference $E_{ij}(\delta) = E_j - E_i$ between the involved DQD eigenstates [see Figure \ref{fig1} (d)], while the rate $\gamma$ summarizes the effect of decoherence on the DQD susceptibility.
Importantly, $\chi_{ij}$ is proportional to the difference of the time-averaged state populations $\Delta p_{ij} = p_i - p_j$.
It can be altered through either temperature \cite{Burkard2016, mi_high_2017} or, as in our experiments, a bias-induced non-equilibrium state population \cite{Viennot2014, mi_high_2017, landig_microwave-cavity-detected_2019}.

\section{Device Details}
%

We measure a BLG DQD in a hybrid cQED sample shown schematically in Figure \ref{fig1}(c), similar to reference \cite{Ruckriegel2024}.
The DQD is dispersively coupled to an on-chip microwave circuit and probed simultaneously with transport measurements.
The sample is mounted in a dilution cryostat with a base temperature of $\SI{10}{\milli\kelvin}$.

The DQD is formed in a van der Waals heterostructure, assembled using the dry transfer technique \cite{Yankowitz2019} from BLG encapsulated in hexagonal boron-nitride (hBN, top: $\SI{27}{\nano\meter}$, bottom: $\SI{37}{\nano\meter}$) with a graphite back gate.
The nanostructure is defined using two layers of metallic top gates, separated by $\SI{20}{\nano\meter}$ of aluminum oxide deposited through atomic layer deposition.
A first layer of top gates (split gates) defines a channel with a lithographic width of $\SI{100}{\nano\meter}$.
The split gates are set to $\SI{4.4}{\volt}$, which, in combination with $\SI{-8}{\volt}$ applied to the back gate, yields a displacement field of $D = \SI{0.65}{\volt/\nano\meter}$ that opens a band gap of $E_\mathrm{gap} \approx \SI{60}{\milli\electronvolt}$ \cite{Icking2022}.
Tuning the Fermi energy to the band gap underneath the split gates enables electrostatic confinement of valence band holes to a one-dimensional  channel.

The second set of top gates (finger gates) are placed across the channel in order to locally form barriers or to confine QDs.
In our experiments, two neighboring finger gates define a DQD [see Figure \ref{fig1} (c)].
Their respective gate voltages $V_\mathrm{L}$ and $V_\mathrm{R}$ control the electrochemical potentials $\mu_\mathrm{L}$ and $\mu_\mathrm{R}$ of the left and right QD.
Voltage values are chosen such that the finger gates accumulate electrons in the p-type channel. 
An additional finger gate with $V_\mathrm{BL}$ [not shown in Figure \ref{fig1} (c)] defines the barrier to the left lead and is also employed to tune the left QD instead of $V_\mathrm{L}$.  
A finite source-drain bias voltage $V_\mathrm{SD}$ (applied symmetrically with respect to measurement ground) probes direct current $I_\mathrm{SD}$ through the structure.
Varying the gate voltages while probing direct transport reveals the charge stability diagram of the DQD in the few-carrier regime (see Figure \ref{fig6} in Appendix \ref{app:csd}).
We identify the charge configurations $(n_\mathrm{L}, n_\mathrm{R})$, where $n_\mathrm{L (R)}$ denotes the electron occupation of the left (right) QD.

One plunger gate of the DQD ($V_\mathrm{L}$) is connected directly to the resonator center conductor and DC-biased through a voltage tap. 
We fabricate the superconducting microwave circuit from a $\SI{25}{\nano\meter}$ thick film of NbTiN with the sheet kinetic inductance $L_\mathrm{kin,\square} = \SI{70}{\pico\henry}$.
The coplanar waveguide resonator is characterized by the resonance frequency $f_\mathrm{r} = \SI{5.241}{\giga\hertz}$.
It is designed with a lithographic width of $\SI{1}{\micro\meter}$, resulting in a characteristic impedance of $\SI{1}{\kilo\ohm}$.
This high impedance enhances the bare dipole coupling strength $g_0$ by a factor $\sqrt{20} \approx 4.5$ compared to a conventional $\SI{50}{\ohm}$ resonator.
The dispersive resonator shift is detected as a change of microwave transmission $S_{21}$ probed at $f_\mathrm{r}$.

\section{Zero-field spin and valley blockade}

\begin{figure}[t!]
\includegraphics{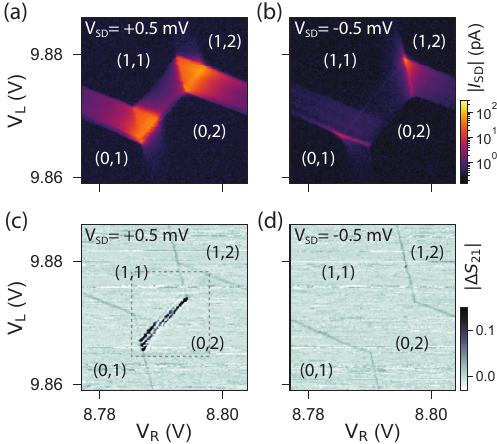}
\caption{\label{fig2} Pauli blockade at the $(1,1) \leftrightarrow (0,2)$ charge transition measured in direct transport and in the microwave response of the resonator.
(a) In the forward direction with $V_\mathrm{SD} > 0$, current is large within the finite bias triangles. 
(b) In the blocked direction with $V_\mathrm{SD} < 0$, current is suppressed by more than two orders of magnitude due to selection rules of the involved ground states.
(c) The microwave response shows three distinct features at the base of the triangles. 
A high-resolution measurement of the outlined region is presented in Figure \ref{fig2} (a). 
(d) The blockade is also detected in the microwave response as the signal from the interdot transition is absent.}
\end{figure}

Spin and valley blockade are important mechanisms for qubit readout in BLG QD qubits, analogous to Pauli spin blockade in conventional spin qubits. 
In the following, we focus on the interdot transitions $(1,n-1) \leftrightarrow (0,n)$ between charge configurations that involve a total of $n=2$ or $n=3$ carriers, i.e., $(1,1) \leftrightarrow (0,2)$ and $(1,2) \leftrightarrow (0,3)$.
Transport measurements with finite positive and negative source-drain bias reveal pronounced Pauli blockade at zero magnetic field, shown for the two-carrier transition in Figs. \ref{fig2} (a) and (b).
With two carriers in the quantum dot, the transition in the direction $(1,1) \leftarrow (0,2)$ shows a large current $I_\mathrm{SD}$ inside the finite bias triangles.
However, in the opposite bias direction $(1,1) \rightarrow (0,2)$, transport is suppressed by more than two orders of magnitude, with the remaining current barely detectable.
Similarly, for three carriers the $(1,2) \rightarrow (0,3)$ transition is blocked (see Figure \ref{fig7} in Appendix \ref{app:3carrier} for more details).
Blocked transport inside the bias window indicates a large maximal spin or valley ground state polarization of the $(1,n-1)$ charge configuration, that is incompatible with the $(0,n)$ ground state \cite{tong_pauli_2023}.
Selection rules prevent tunneling from the $(1,n-1)$ to the $(0,n)$ charge configuration, once a highly polarized state is loaded in the blocked bias direction.

More specifically, in the case of two carriers, the set of degenerate $(1,1)$ ground states contains a maximally valley- and spin-polarized valley-triplet spin-triplet state $T_\mathrm{v}T_\mathrm{s}$ with total valley and spin quantum numbers $\tau=1$ and $\sigma=1$. 
The Pauli exclusion principle blocks the charge transition to the $(0,2)$ ground state, a valley-singlet spin-triplet ($\tau=0, \sigma=1$), since $\tau_\mathrm{(1,1)} > \tau_\mathrm{(0,2)}$. 
As the valley d.o.f. causes the blockade, we speak of \emph{valley blockade} for the two-carrier transition.
The three-carrier transition features \emph{spin blockade}: tunneling from the maximally spin-polarized $(1,2)$ ground state with $\sigma=3/2$ to any of the $(0,3)$ ground states with $\sigma = 1/2$ is forbidden.
These insights are generally limited to ground state transport, as the measurements do not disclose information about the spin or valley polarization of excited states.
Instead, characteristics of excited states are inferred from magnetic field dependence of transport features.
However, at zero magnetic field, the magnitude of energy splittings between DQD states, that in BLG arise from Kane-Mele SO interaction \cite{Kane2005}, are not resolved in typical transport measurements such as the data shown in Fig.~\ref{fig2} (a) and (b).

The bias-dependent effect of Pauli blockade is also apparent in the microwave transmission $|S_{21}|$ [see Figs. \ref{fig2} (c) and (d)].
In contrast to direct transport, the resonator's response to the interdot charge transition under positive bias reveals a set of distinct resonances.
Three sharp, tightly spaced lines emerge in the direction of constant detuning for the positive bias, shown in panel (c).
This signature arises from virtual dipole transitions between ground and excited states of the two involved charge configurations.
The response vanishes for the blocked transport direction as a result of Pauli blockade: once the system is locked in the blocked state, selection rules suppress the transition dipole moment that gives rise to the dispersive resonator shift \cite{landig_microwave-cavity-detected_2019}.

\section{Finite bias resonator response}
\label{sec:RF}
%

\begin{figure*}[t!]
\includegraphics{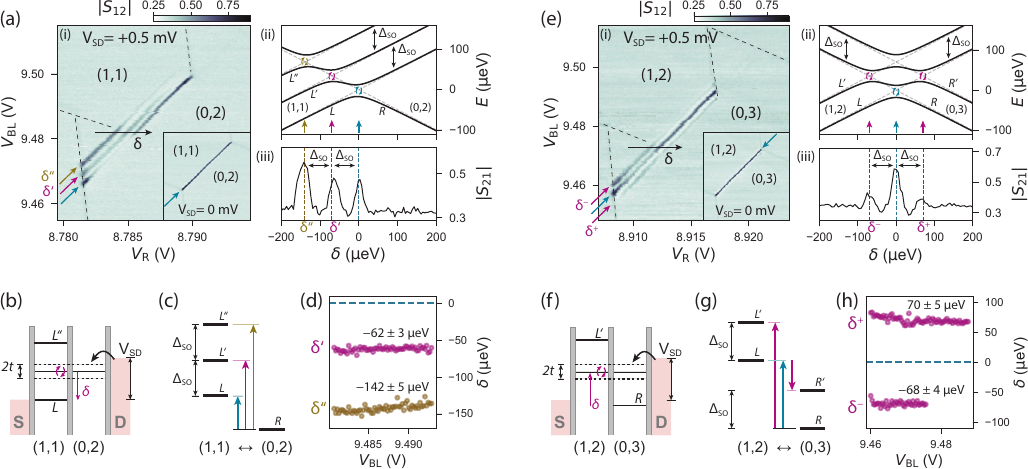}
\caption{\label{fig3} 
Finite-bias resonator response at the $(1,1) \leftrightarrow (0,2)$ [panels (a) - (d)] and $(1,2) \leftrightarrow (0,3)$ [panels (e) - (h)] transitions.
The microwave transmission $S_{21}$ at $V_\mathrm{SD} = +\SI{0.5}{\milli\volt}$ shows distinct features at finite DQD detunings.
(a) The $n=2$ resonator signal is explained qualitatively by two low-lying excited states of the $(1,1)$ charge configuration, labeled as $L^\prime$ and $L^{\prime \prime}$.
(ii) Detuning dependence of the two-carrier DQD states with finite tunnel coupling. 
Panel (iii) shows a line cut along the detuning axis marked in (i)
(b) Finite source-drain bias creates a non-equilibrium population of excited states if the hybridized charge states (dashed lines) fall within the bias window. Visibility of the excited state transitions is therefore enhanced inside the lower bias triangle.
(c) Transitions at finite detuning energies indicated by teal, purple and yellow correspond to the arrows of same color in (a).
(d) Detuning energies of excited state transitions extracted from fits to the signal peaks.
(e) - (h) The $n=3$ transition similarly shows three distinct features in $S_{21}$.
However, now the central feature (teal arrow) corresponds to the $L \leftrightarrow R$ ground-state transition, as evident from its high visibility in the region between lower and upper triangle.
The observed signal is explained by the presence of one low-lying excited state in either charge configuration, i.e. $L^\prime$ and $R^\prime$.
}
\end{figure*}

We take a closer look at the finite-bias resonator response for two and three carriers.
Figure \ref{fig3} presents the microwave transmission $|S_{21}|$ measured at the resonance frequency $f_\mathrm{r}$ for $V_\mathrm{SD} = +\SI{0.5}{\milli\volt}$ at the interdot transitions with $n=2$ (a) and $n=3$ (e).
In contrast to the zero-bias response [insets in panels (i)], both finite-bias measurements reveal three sharp resonances at different DQD energy detunings $\delta = \mu_\mathrm{L} - \mu_\mathrm{R}$, marked by arrows in (i).
We convert the gate-voltage $V_\mathrm{R}$ to an energy detuning $\delta$ using the gate lever arm $\alpha = 0.09$ (extracted from finite bias measurements). 

The additional finite-bias resonances arise from virtual dipole transitions involving excited states that become populated under finite bias conditions.
Panels (ii) in Figs. \ref{fig3} (a) for $n=2$ and (e) for $n=3$ show the DQD energy levels as a function of $\delta$.
The level spectra give a qualitative understanding of how additional peaks in resonator transmission at finite $\delta$ arise from virtual dipole transitions involving excited states at the indicated avoided crossings.   
Their visibility is enhanced compared to the situation for $V_\mathrm{SD} = \SI{0}{\milli\volt}$ (see insets).
A source-drain bias in the forward direction introduces a non-equilibrium state population, increasing the population difference of the involved states that couple to the microwave resonator.
Consequently, the relative strength of the transitions changes inside the bias triangles outlined by dashed lines in Panels (i) of (a) and (e).

For the $n=2$ transition in Figure $\ref{fig3}$ (a), panel (iii) shows a line cut of the data along the energy detuning axis $\delta$ indicated by the black arrow in (i).
The two additional features at negative detuning stem from two excited states of the $(1,1)$ configuration.
We label them as $L^\prime$ and $L^{\prime \prime}$ in (ii).
The excited states lead to interdot charge transitions $L^\prime \leftrightarrow R$ and $L^{\prime \prime} \leftrightarrow R$ at negative detunings $\delta^\prime$ and $\delta^{\prime\prime}$, in addition to the ground-state transition $L \leftrightarrow R$ observed already at zero bias that defines $\delta = 0$.
The line cut in (iii) therefore represents a direct measurement of the excited state energy splittings from the detuning difference between individual dipole transitions.
We will see in Section \ref{sec:states} that this energy splitting corresponds to the intrinsic SO gap $\Delta_\mathrm{SO}$ of BLG QDs.

Figure $\ref{fig3}$ (b) and (c) show a schematic representation of the $n=2$ DQD states that lead to the observed resonator transmission. 
The tunnel coupling $t$ hybridizes the respective $(1,1)$ and $(0,2)$ states when they align in energy for $\delta = 0$ (teal), $\delta^\prime = - \Delta_\mathrm{SO}$ (purple) and $\delta^{\prime \prime} = - 2 \Delta_\mathrm{SO}$ (yellow), resulting in a dispersive features at those detuning values.
The signal strength of these excited state transitions depends strongly on the applied source-drain bias:
they are most visible if the hybridized states lie within the bias window, i.e. inside the lower bias triangle.
Tunneling-in from the leads to the hybridized states of the two-level system results in a non-equilibrium state occupation.
Population of higher DQD states leads to a finite population difference $\Delta p_{ij}$ for the $i \leftrightarrow j$ dipole transition, even when the state $i$ is not the system's ground state.
For the zero-bias measurement, i.e. with $V_\mathrm{SD} = \SI{0}{\milli\volt}$ [shown in the inset of Figure \ref{fig3} (a)], occupation of excited states is only due to thermal population which is strongly suppressed because $\Delta E \gg k_\mathrm{B}T$. 
As a result, the ground-state transition $L \leftrightarrow R$ dominates (marked by teal arrow) and no other resonances are visible.
The situation \emph{between} the bias triangles resembles the situation with zero bias: as the hybridized states lie outside the bias window, the state population is not affected by the bias and excited state transitions are less pronounced.
The remaining signal therefore stems predominantly from the ground-state transition at $\delta = 0$.
 
For the $n=3$ charge transition shown in Figure $\ref{fig3}$ (e), three features are also apparent inside the lower bias triangle.
In contrast to the situation with two carriers, the ground-state transition ($\delta=0$) corresponds to the central feature marked by the teal arrow, as it remains most pronounced between the triangles, while the other two features are weak.
The two excited state dipole transitions at finite positive and negative detuning inside the lower bias triangle stem from the first excited state in either one of the involved charge configurations $(1,2)$ and $(0,3)$.
We label the respective excited states as $L^\prime$ and $R^\prime$ in (ii). 
Consequently, their dipole transitions due to a finite tunnel coupling to the respective ground states $L$ and $R$ occur at positive detuning $\delta^+$ and negative detuning $\delta^-$ [marked by purple arrows in (e)].
The signal therefore represents a direct measurement of the low-energy excited states of the $(1,2)$ and $(0,3)$ charge configurations.

In order to quantify the energy gaps in the two- and three-carrier spectra due to SO coupling, we extract the detuning energies $\delta^\prime$ and $\delta^{\prime\prime}$ from the measurement in Figure \ref{fig3} (a), as well as $\delta^+$ and $\delta^{-}$ from (e). 
We fit Gaussians to the three signal peaks for a range of $V_\mathrm{BL}$.
The resulting center positions of the peaks relative to $\delta = 0$ (teal) are shown in panel (d) and (h).
Average values within the measured range of $V_\mathrm{BL}$ are $\delta^\prime = \SI{-62\pm3}{\micro\electronvolt}$ and $\delta^{\prime\prime} = \SI{-142\pm5}{\micro\electronvolt}$ for the two-carrier transition and $\delta^+ = \SI{70\pm5}{\micro\electronvolt}$ and $\delta^{-} = \SI{-68\pm4}{\micro\electronvolt}$ for the three-carrier transition.

\section{Few-carrier states in a BLG DQD}
\label{sec:states}
\begin{figure*}[t!]
\includegraphics{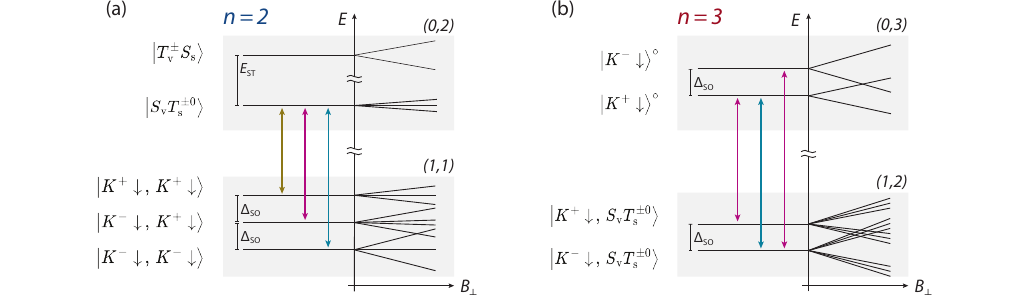}
\caption{\label{fig4} Schematic of interdot transitions between $(1,n-1)$ and $(0,n)$ charge configurations for $n=2,3$.
(a) Two-carrier states in a DQD. 
With negligible exchange interaction for the $(1,1)$ charge configuration, the 16 possible spin-valley states split into three bundles, separated in energy by $\Delta_\mathrm{SO}$. 
Their coupling to the $(0,2)$ ground state, a valley-singlet spin-triplet, gives rise to the dispersive signature presented in Figure \ref{fig3} (a).
(b) Three-carrier interdot transition.
Both the $(1,2)$ and $(0,3)$ charge configuration feature one excited state, separated from the ground-state by $\Delta_\mathrm{SO}$.
In contrast to $n=1$, the corresponding excited-state transitions (purple arrows) are not blocked and lead to dispersive resonator interaction at finite positive and negative detuning values [see Figure \ref{fig3} (e)].}
\end{figure*}

%
We now turn to a more detailed discussion of the states involved in the few-carrier interdot charge transitions. 
An overview of states for the relevant charge configurations is given in Figure \ref{fig4}, highlighting the dipole transitions for $n=2,3$.
As a omparison, we discuss the single-carrier DQD interdot transition between $(1,0) \leftrightarrow (0,1)$ in Appendix \ref{app:n1}.

States of $n$ carriers in a \emph{single} quantum dot, i.e. $(0,n)$  in the DQD notation, are characterized by their total valley and spin quantum numbers $\tau$ and $\sigma$, respectively.
One carrier in the right dot, i.e. $(0,1)$, forms Kramers doublets $\ket{K^- \downarrow}_{(0,1)}$ and $\ket{K^+ \uparrow}_{(0,1)}$ with parallel, or $\ket{K^+ \downarrow}_{(0,1)}$ and $\ket{K^- \uparrow}_{(0,1)}$ with anti-parallel alignment of valley and spin magnetic moments.
At zero magnetic-field, the pairs are split in energy by Kane-Mele SO interaction of strength $\Delta_\mathrm{SO}$ \cite{Kurzmann2021, Banszerus2021, Duprez2024}. A perpendicular magnetic field lifts the degeneracies with a spin $g$-factor $g_\mathrm{s} = 2$ and a significantly larger valley $g$-factor $g_\mathrm{v} \approx 10 - 20$ \cite{tong_tunable_2020}.

For two carriers ($n=2$), spins and valleys pair up as a valley-singlet spin-triplet ground state $\ket{S_\mathrm{v} T_\mathrm{s}^{\pm,0}}_{(0,2)}$ with $\tau=0$ and $\sigma=1$ that is threefold degenerate at zero magnetic-field (with spin magnetic moment $m_\sigma = -1, 0, 1$). 
This spin-1 ground-state has been consistently observed in experiments \cite{kurzmann_excited_2019, Moeller2021, tong_pauli_2022, Moeller2024}, explained in theory \cite{Knothe2020, Knothe2022}, and is further corroborated by the strong zero-field Pauli blockade presented above.
The first excited state is a valley-triplet spin-singlet state $\ket{T_\mathrm{v}^{\pm} S_\mathrm{s}}_{(0,2)}$ \cite{tong_pauli_2022, Moeller2021, Moeller2024} that we measure at $E_\mathrm{ST} \approx \SI{0.9}{\milli\electronvolt}$ higher in energy.

Three carriers ($n=3$) combine their valley and spin degrees in one of four different ways that result from \emph{removing} a single carrier with arbitrary valley and spin from a fully occupied shell of four carriers \cite{Tong2024}.
The corresponding state spectrum therefore resembles the single-carrier case, but inverted in energy [see Figure \ref{fig4} (b)].
Four three-carrier states form two doublets split by $\Delta_\mathrm{SO}$ at zero magnetic field. 
For convenience, we label them with valley and spin of the carrier removed from the full shell (symbolized by $^\circ$).
The doublet of $\ket{K^+ \downarrow}^\circ_{(0,3)}$ and $\ket{K^- \uparrow}^\circ_{(0,3)}$ is lowered in energy, while $\ket{K^- \downarrow}^\circ_{(0,3)}$ and $\ket{K^+ \uparrow}^\circ_{(0,3)}$ are lifted.

Adding one carrier to the neighboring quantum dot extends the excited state spectrum.
We assume that, to first order, the two systems can be seen as independent so that the state energies of $(1, n)$ are the sum of the energies of $(1,0)$ and $(0,n-1)$ \cite{tong_pauli_2022, Tong2024, Moeller2024}.
In effect, the $(1,1)$ charge configuration represents the combination of two single-carrier spectra, each with a SO gap at zero magnetic-field.
The result is a fourfold degenerate ground state with a combination of any of the spin-valley-aligned lower Kramer's pair states in the left or right QD.
Two of these states, namely $\ket{K^- \downarrow, K^- \downarrow}_{(1,1)}$ and $\ket{K^+ \uparrow, K^+ \uparrow}_{(1,1)}$, have total valley and spin quantum numbers $\tau = 1$ and $\sigma = 1$.
Their valley-polarization gives rise to the valley blockade in the $(1,1) \rightarrow (0,2)$ bias direction:
they are prohibited from tunneling to the $\tau=0$ $(0,2)$ ground state and have a vanishing transition dipole matrix element with this state.
The first excited state (split off by $\Delta_\mathrm{SO}$) comprises all eight combinations for which the SO interaction energies on left and right dot cancel.
A second excited state is again fourfold degenerate with only anti-parallel spin-valley pairings and lies at $2\Delta_\mathrm{SO}$ relative to the ground-state.

The $(1,2)$ charge configuration features ground and excited states split by $\Delta_\mathrm{SO}$ that are each six-fold degenerate.
They are a combination of the two-carrier spin-triplet ground state in the right QD and the Kramer's doublet with parallel or anti-parallel spin-valley alignment in the left QD. 
This combination of three carriers results in a subset of maximally spin-polarized ground states with $\sigma = 3/2$ that are blocked from tunneling to the $(0,3)$ ground state with $\sigma = 1/2$, explaining the observed spin blockade at zero field (see Figure \ref{fig5}).

With this discussion, we described the nature of the ground- and low-lying excited states that are detected by the microwave resonator at the $n=2$ and $n=3$ interdot transitions. 
The two low-lying excited states for the $(1,1)$ charge configuration, labeled in Section \ref{sec:RF} as $L^\prime$ and $L^{\prime\prime}$ [see Fig.~\ref{fig3} (a, ii)], correspond to the eight-fold and fourfold degenerate states at energies $\Delta_\mathrm{SO}$ and $2\Delta_\mathrm{SO}$ above the ground state.
The $R$ ground state of $(0,2)$ is the two-carrier spin-triplet.
Together, they give rise to the three tightly spaced resonances that are seen in Figs.~\ref{fig3} (a, i \& iii) and discussed above. 
On the other hand, both the $(1,2)$ and the $(0,3)$ charge configurations have one low-lying excited state [$L^\prime$ and $R^\prime$ in Fig.~\ref{fig3} (e, ii)].
They are each split from their respective ground states by $\Delta_\mathrm{SO}$ due to the effect of SO interaction that affects Kramers pairs at zero field. As a result we observe the three resonances in Fig.~\ref{fig3} (e, i \& iii).

\section{Discussion}
%
In Section \ref{sec:RF} we have extracted the energy gaps between ground states and low-lying two- and three-carrier excited states.
The peaks are well resolved in the resonator signal, see Figure \ref{fig3} (a) and (e), allowing for a precise quantitative determination of the respective energy gaps with standard deviations of $\SI{5}{\micro\electronvolt}$ and less.
The full-width-at-half-maximum (FWHM) of the signal peaks is within a range of $15-\SI{30}{\micro\electronvolt}$, corresponding roughly to the energy splitting $2t/h \approx \SI{6}{\giga\hertz}$ of the hybridized DQD states. 
The advantage of the high energy resolution achieved with our measurement technique becomes apparent with measurements in a small external magnetic field.
As we describe in Appendix \ref{app:magnet}, we apply a perpendicular field of $B_\perp = \SI{50}{\milli\tesla}$ to the sample, lifting state degeneracies at the $(1,1) \leftrightarrow (0,2)$ interdot transition.
The resonator signal corresponding to the $\delta=0$ ground-state virtual dipole transitions [teal in Figure \ref{fig3} (a)] splits into two clearly distinguishable peaks as a result of the Zeeman effect.
Those signal peaks are separated in energy by $\Delta E \approx \SI{30}{\micro\electronvolt}$.
Energy splittings of this magnitude are not detectable in typical DQD transport  \cite{Banszerus2021, tong_pauli_2022, banszerus_particlehole_2023, Tong2024, Moeller2024}, as inelastic tunneling processes mask the resonant processes through low-lying excitations.

Furthermore, is evident from the microwave-detected Pauli blockade discussed above, that the resonator signal is selective to the transition's valley and spin states. 
The microwave signal reflects the presence or absence of an interdot dipole moment which is indicative of level hybridization.
Dispersive interaction can therefore be employed for fast state readout \cite{zheng_rapid_2019, landig_microwave-cavity-detected_2019, Ruckriegel2024}.
Consider, for example, the ground-state transition for $n=2$:
out of the four degenerate $(1,1)$ ground-states, the two states $\ket{K^- \downarrow, K^- \downarrow}_{(1,1)}$ and $\ket{K^+ \uparrow, K^+ \uparrow}_{(1,1)}$ have total valley quantum number $\tau = 1$ and are therefore blocked from tunneling to any of the $(0,2)$ valley-singlet ground states.
Only the two states with a net-zero valley- and spin-polarization (i.e. $\ket{K^- \downarrow , K^+ \uparrow}_{(1,1)}$ and $\ket{K^+ \uparrow, K^- \downarrow}_{(1,1)}$) couple to one of the $(0,2)$ states, namely $\ket{S_\mathrm{v} T_\mathrm{s}^0 }_{(0,2)}$.
This sub-space of the tunneling Hamiltonian can be written as
\begin{equation}
    H_{\sigma, \tau = 0}
    \begin{pmatrix}
        0 & 0 & -t\\
        0 & 0 & t\\
        -t^* & t^* & \delta
    \end{pmatrix} , 
\end{equation}
in the basis ($\ket{K^- \downarrow , K^+ \uparrow}_{(1,1)}$, $\ket{K^+ \uparrow, K^- \downarrow}_{(1,1)}$, $\ket{S_\mathrm{v} T_\mathrm{s}^0 }_{(0,2)}$), where $t$ is the tunnel coupling and $t^*$ its complex conjugate.
The $(1,1)$ states can in turn be rewritten as symmetric and anti-symmetric superpositions of Kramers pairs with zero valley or spin, i.e. as a \emph{Kramers singlet} and \emph{triplet}:
\begin{align}
    \ket{A}_{(1,1)} = \frac{1}{\sqrt{2}}\left( \ket{K^- \downarrow, K^+ \uparrow} - \ket{K^+ \uparrow, K^- \downarrow} \right) \\
    \ket{B_0}_{(1,1)} = \frac{1}{\sqrt{2}}\left( \ket{K^- \downarrow, K^+ \uparrow} + \ket{K^+ \uparrow, K^- \downarrow} \right)  
\end{align}
Of the two states, only the anti-symmetric Kramer's singlet $\ket{A}$ can couple to $\ket{S_\mathrm{v} T_\mathrm{s}^0}_{(0,2)}$ via interdot tunneling, while the symmetric triplet $\ket{B_0}$ does not.
The resulting Hamiltonian in this new basis reads
\begin{equation}
    H_{\sigma, \tau = 0}^{B_0/A}
    \begin{pmatrix}
        0 & 0 & 0\\
        0 & 0 & - \sqrt{2}t\\
        0 & -\sqrt{2}t^* & \delta
    \end{pmatrix} .
\end{equation}
It is similar to its spin-only counterpart in other qubit systems that forms the basis of singlet-triplet qubits \cite{Petta2005}.
The dynamics of a Kramers singlet-triplet qubit can be described by the Hamiltonian
\begin{equation}
    H = \frac{1}{2} J(\delta) \sigma_z + \frac{1}{2} \Delta g_\mathrm{v} \mu_\mathrm{B} B_\perp \sigma_x.
\end{equation}
All electric driving of rotations about both qubit axes can be achieved by a detuning-dependent spin-valley exchange $J(\delta)$, and by electrostatic modulation of the valley g-factor difference $\Delta g_\mathrm{v}$ \cite{tong_tunable_2020} in small perpendicular magnetic field $B_\perp$. Alternatively, magnetic driving with a time-modulated perpendicular magnetic field $B_\perp(t)$ would also be possible.

\begin{figure}[t!]
\includegraphics{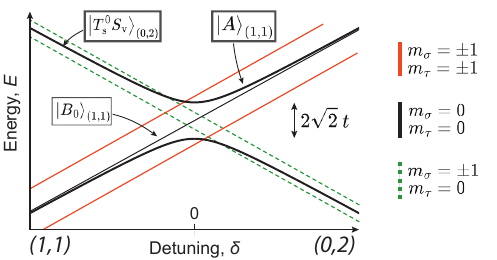}
\caption{\label{fig5} State-selective dipole coupling of a Kramer's singlet-triplet qubit.
Energy dispersion of the $(1,1) \leftrightarrow (0,2)$ interdot transition as a function of detuning energy $\delta$.
A small perpendicular magnetic field splits spin- and valley polarized states off the unpolarized state (teal and green lines).
The valley-singlet spin-triplet $(0,2)$ ground state $\ket{S_\mathrm{v} T^0_\mathrm{s}}_{(0,2)}$ couples only to the anti-symmetric Kramers-singlet $\ket{A}_{(1,1)}$, resulting in a finite transition dipole sensed by the on-chip microwave resonator.
The symmetric Kramers-triplet $\ket{B_0}_{(1,1)}$ on the other hand does not couple.}
\end{figure}

As the resonator signal is sensitive to dipole transitions between charge states $(1,1) \leftrightarrow (0,2)$, the dispersive readout can distinguish between Kramer's singlet $\ket{A}_{(1,1)}$ and triplet $\ket{B_0}_{(1,1)}$.
Figure \ref{fig5} schematically depicts the ground-state to ground-state transition plotted for a finite magnetic field. 
Only the $\ket{A}_{(1,1)}$ state couples with strength $\sqrt{2} t$ to $\ket{S_\mathrm{v} T^0_\mathrm{s}}_{(0,2)}$.
The singlet-hybridization with an avoided crossing of $2\sqrt{2}t$ is therefore detected by the microwave resonator, whereas the transition to the triplet $\ket{B_0}_{(1,1)}$ exhibits a vanishing dipole moment.
This could serve as a fast readout scheme similar to reference \cite{zheng_rapid_2019}.
When either $\ket{A}_{(1,1)}$ or $\ket{B_0}_{(1,1)}$ are loaded in the $(1,1)$ configuration, the resonator signal at $\delta=0$ differentiates between the Kramer's singlet (dispersive shift) and triplet (no shift).
The readout time is fundamentally limited by the resonator linewidth of $\kappa / 2\pi \approx \SI{2.2}{\mega\hertz}$. 

\section{Conclusion \& Outlook}
%
We employed a high-impedance microwave resonator to investigate interdot charge transitions with either two or three electrons in a BLG DQD.
In both cases, we measured strong Pauli blockade at zero magnetic field that arises from spin- or valley-polarized ground states.
Dipole coupling with the resonator produces a bias-dependent microwave signature: 
while the blocked bias-direction results in a vanishing transition dipole, we observe distinct features in the resonator response in the forward bias direction.
They arise from dipole transitions involving higher excited states, whose visibility is enhanced by a non-equilibrium state population under finite bias.
Our dispersive measurement of dipole transitions demonstrates an exceptional energy resolution compared to direct transport through DQDs that relies on sequential tunneling.
We take advantage of the sensitivity of cQED measurements and probe energy gaps between low-lying states of two- and three-carrier charge configurations at zero magnetic field. 
They arise from the Kane-Mele SO interaction intrinsic to BLG that we quantify for each of the interdot transitions.
The technique can be more generally employed to investigate spin-valley physics and mixing mechanisms \cite{borjans_probing_2021} at other interdot transitions of BLG QDs, e.g. the particle-hole symmetric single-carrier transition \cite{banszerus_particlehole_2023} or transitions with more carriers \cite{tong_pauli_2023}.
Investigating these few-carrier valley and spin states and their mutual couplings at low magnetic fields is crucial for evaluating their use as qubit states in BLG.
The bias-dependent resonator response highlights how dipole coupling to the DQD can serve as dispersive readout that is both state-selective and fast \cite{zheng_rapid_2019, Ruckriegel2024}.
We outline how the presented measurements discriminate between states of a possible qubit implementation in a BLG DQD, the Kramers singlet-triplet qubit.
%
%
%
%
%
%
%
%
%
%
\section*{Acknowledgements}
We are grateful for P. M\"arki and T. B\"ahler, the FIRST staff for their technical support, as well as Alexander Flasby for their assistance with NbTiN deposition.
We thank Pasquale Scarlino for helpful discussion.
We acknowledge financial support from the European Graphene Flagship, the ERC Synergy Grant Quantropy, and the European Union’s Horizon 2020 research and innovation program under grant agreement number 862660/QUANTUM E LEAPS and NCCR QSIT (Swiss National Science Foundation). K.W. and T.T. acknowledge support from the JSPS KAKENHI (Grant Numbers 21H05233 and 23H02052) , the CREST (JPMJCR24A5), JST and World Premier International Research Center Initiative (WPI), MEXT, Japan.

M.J.R., T.I. and K.E. conceived the experiment. 
M.J.R. fabricated the device with inputs from D.K.
The NbTiN film was deposited by M.B.P.
M.J.R. performed the measurements with assistance from C.A. and R.B.
M.J.R. analyzed and interpreted the data with inputs from C.A., C.T., D.K., A.O.D., T.I. and K.E. 
K.W. and T.T. provided the hBN crystals.
M.J.R. wrote the manuscript with contributions from all authors.

\appendix

\section{Few-carrier charge stability diagram}
\label{app:csd}
\begin{figure}[h!]
\includegraphics{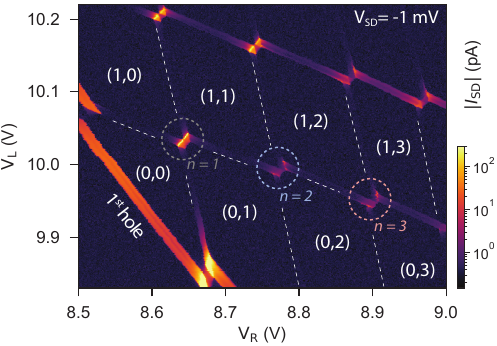}
\caption{\label{fig6} Charge stability diagram of the DQD. 
Source drain current $I_\mathrm{SD}$ measured with $V_\mathrm{SD} = \SI{-1}{\milli\volt}$ as a function of left and right plunger gate voltages.
The discussion above focuses on the interdot charge transitions with a total of $n=1,2,3$ electrons, highlighted with circles.}
\end{figure}

The charge stability diagram of the DQD is shown in Figure \ref{fig6}.
We measure the source drain current $I_\mathrm{SD}$ at a bias of $V_\mathrm{SD} = \SI{-1}{\milli\volt}$ while varying the left and right plunger gate voltages.
Charging lines indicate the addition of the first electrons in the left or right QD that are accumulated by more and more positive gate voltages.
The charge configurations $(n_\mathrm{L}, n_\mathrm{R})$ are labeled accordingly.
For lower $V_\mathrm{L}$ and $V_\mathrm{R}$, a hole QD forms between the gates, as seen from the diagonal line of large current.

\section{Three-carrier zero-field Pauli blockade}
\label{app:3carrier}

\begin{figure}[h!]
\includegraphics{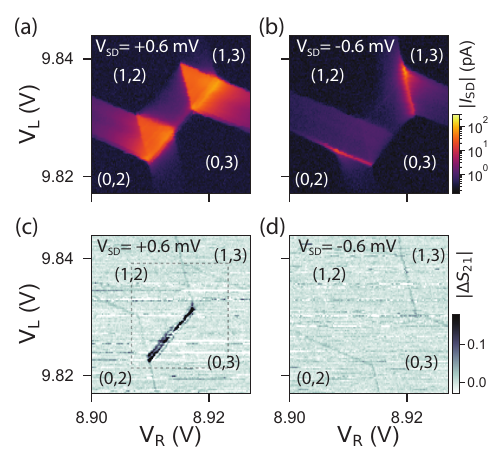}
\caption{\label{fig7} Pauli blockade at the $(1,2) \leftrightarrow (0,3)$ charge transition measured in direct transport and in the microwave response of the resonator.
(a) In the forward direction with $V_\mathrm{SD} > 0$, current is large within the finite bias triangles. 
(b) In the blocked direction with $V_\mathrm{SD} < 0$, current is strongly suppressed due to Pauli spin blockade.
(c) The microwave response in the forward bias direction shows three features at the base of the triangles.
See Figure \ref{fig3} (d) for high-resolution data of the marked rectangle in (c).
(d) Blockade is also evident in the resonator response as the signal from the interdot transition is absent.}
\end{figure}

Figure \ref{fig7} presents the blockade data for the three-carrier interdot transition $(1,2) \leftrightarrow (0,3)$.
As in the two-carrier case, both direct transport [(a) and (b)] and dispersive resonator measurements [(c) and (d)] show pronounced Pauli blockade.
Under forward (positive) bias, the bias triangles are complete.
Ground-state transport is permitted between in the $(1,2) \leftarrow (0,3)$ direction resulting in large current inside the bias window.
Any of the three-carrier states that may be loaded couple to at least one of the $(1,2)$ ground states.
In contrast, two out of the six degenerate $(1,2)$ ground states are spin-polarized with $\sigma = 3/2$, blocking them from tunneling to any of the $(0,3)$ states with $\sigma = 1/2$. 
The resonator response for positive $V_\mathrm{SD}$ shows a set of features close to zero energy detuning $\delta$. 
They are a signature of interdot transitions with finite tunnel coupling, including excited states of the $(1,2)$ and $(0,3)$ charge configurations.
A measurement with higher resolution is presented in Figure \ref{fig3} (e) and discussed in Section \ref{sec:RF}.
Pauli blockade in the $(1,2) \rightarrow (0,3)$ direction suppresses tunneling and the transition dipole moment such that the dispersive resonator signal vanishes.
For more details on three-carrier states and Pauli blockade see references \cite{tong_pauli_2023, Tong2024}.

\section{Comparison to single-carrier interdot transition}
\label{app:n1}

\begin{figure}[h!]
\includegraphics{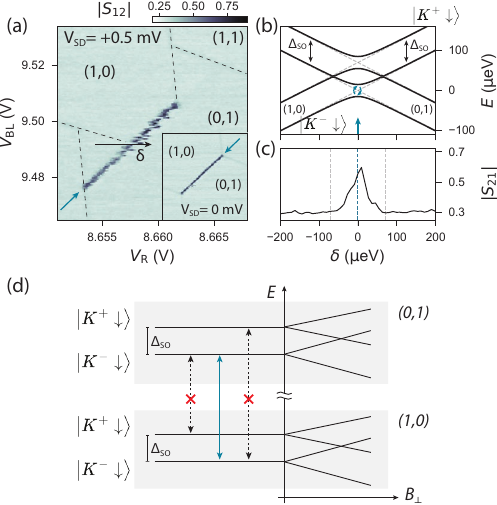}
\caption{\label{fig8} Finite bias resonator response at the $(1,0) \leftrightarrow (0,1)$ interdot transition.
(a) Microwave transmission $S_{21}$ for positive bias shows a single feature from the ground-state to ground-state transition, indicated by the teal arrow.
Inset: resonator response for zero source-drain bias.
(b) Energy of DQD states as a function of detuning.
Ground and excited states correspond to Kramers pairs with either parallel or anti-parallel spin-valley alignment.
(c) Line cut along $\delta$ as indicated in (a). 
Only the anti-crossing at $\delta = 0$ leads to dispersive interaction with the resonator.
(d) The states of a single electron in a QD comprise two Kramers pairs split by $\Delta_\mathrm{SO}$ at zero magnetic field.
Only the ground-state to ground-state transition is permitted, resulting in a single dispersive feature.
}
\end{figure}

We also investigated the $(1,0) \leftrightarrow (0,1)$ interdot transition of a single electron in a DQD.
As the charge configurations are symmetric, there is no Pauli blockade.
Figure \ref{fig8} shows the finite bias resonator response.
In contrast to the $n=2$ and $n=3$ interdot transitions described in Section \ref{sec:RF} and shown in Figure \ref{fig3}, only the ground-state to ground-state transition (teal arrow) is detected by the microwave resonator. 
No signal peaks at finite detuning are visible.

While the single-carrier ground state is a Kramers pair with parallel spin and valley (i.e. $\ket{K^- \downarrow}$, $\ket{K^+ \uparrow}$), the excited state is a Kramers pair with anti-parallel alignment ($\ket{K^+ \downarrow}$, $\ket{K^- \uparrow}$).
Notably, the excited state is orthogonal to the ground state.
Assuming that both valley and spin quantum numbers are conserved during tunneling, the transition is blocked, as it would require either a spin or valley flip.
The $n=1$ transition therefore further shows how the dispersive readout is state-selective.
See Figure \ref{fig8} (d) for a schematic representation of the interdot transitions.
The dashed lines indicate the \emph{blocked} transitions: $\ket{K^+ \downarrow}_{(1,0)} \leftrightarrow \ket{K^- \downarrow}_{(0,1)}$ at negative energy detunining and $\ket{K^- \downarrow}_{(1,0)} \leftrightarrow \ket{K^+ \downarrow}_{(0,1)}$ at positive detuning.
Indeed, the finite-bias experimental data in Figure \ref{fig8} (a) and the line cut in (c) show that these transitions carry no dipole moment.
The absence of additional signal peaks at finite DQD detuning $\delta$ indicates that mechanisms that mix the single-carrier ground and excited states are weak.
We conclude that spin and valley flips are extremely rare processes on the time-scale of interdot tunneling, in line with previous observations \cite{garreis_long-lived_2023, Denisov2025}.

\section{Two-carrier transition at finite magnetic field}
\label{app:magnet}

\begin{figure}[h!]
\includegraphics{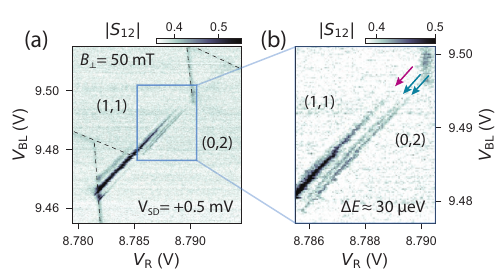}
\caption{\label{fig9} Resonator signal at $B_\perp = \SI{50}{\milli\tesla}$ for the $(1,1) \leftrightarrow (0,2)$ charge transition measured with $V_\mathrm{SD} = \SI{0.5}{\milli\volt}$ forward bias.
The base line at $\delta = 0$ splits in the region between the lower and upper bias triangles [compare to Figure \ref{fig3} (a) for $B_\perp = 0$].
The two ground state signal peaks, marked with teal arrows, are separated by $\Delta E \approx \SI{30}{\micro\electronvolt}$.
}
\end{figure}

We perform measurements in a magnetic field $B_\perp = \SI{50}{\milli\tesla}$ perpendicular to the plane of BLG.
These measurements are made possible by the magnetic field resilience of NbTiN high-impedance resonators with a $w = \SI{1}{\micro\meter}$ narrow center conductor \cite{samkharadze_high_2016}.
The on-chip resonator employed in the presented device maintains a high quality factor at moderate fields.
Figure \ref{fig9} shows the microwave response of the resonator at the $n=2$ interdot transition at $V_\mathrm{SD} = \SI{0.5}{\milli\volt}$ forward bias.
Compared to the measurement at zero field [see Figure \ref{fig3} (a)], the strength of three signal peaks is altered. 
Most notably, the base line at $\delta=0$ in the section between the bias triangles exhibits a small but clearly discernible energy splitting of $\Delta E \approx \SI{30}{\micro\electronvolt}$ [teal arrows in (b)].

\section{Electric dipole coupling}
\label{app:coupling}

The resonator is coupled to a microwave feedline in a hanger type geometry.
The microwave transmission $S_{21}$ through the feedline probes the coupled DQD-resonator system.
It is given by
\begin{equation}
    S_{21} = 1 + \frac{i \kappa_\mathrm{ext}}{2 \Delta_\mathrm{r} - i \kappa + 2 \sum_{i=0}^2 g_{i,i+1} \chi_{i,i+1}}.
\end{equation}
The resonator-probe detuning is $\Delta_\mathrm{r} = 2\pi (f_\mathrm{r} - f)$.
The total resonator linewidth $\kappa = \kappa_\mathrm{int} + \kappa_\mathrm{ext}$ combines internal and external microwave losses.
The last term in the denominator sums over dipole transitions between states $i \leftrightarrow i+1$ of the DQD.
The transition's effective dipole coupling strength $g_{i,j} = g_0 \cdot d_{i,j}$ is proportional to the bare coupling strength $g_0$ and the transition dipole moment $d_{i,j}$, which in turn depends on energy detuning $\delta$ and tunnel coupling $t$.
It is largest when DQD states hybridize as a result of a finite tunnel coupling.
For a simple charge qubit without excited states the effective coupling is $g_{01} = g_0 \cdot 2t / \sqrt{\delta^2 + 4 t^2}$.

The DQD electric susceptibility is given by
\begin{equation}
    \chi_{i,j} = g_{i,j} \cdot \frac{\Delta p_{i,j}}{2\pi f_\mathrm{r} - E_{i,j}/\hbar + i \gamma}.
\end{equation}
Besides the effective dipole coupling $g_{i,j}$, the susceptibility is a function of the energy difference $E_{i,j} = E_j - E_i$ between DQD eigenstates $i$ and $j$.
The total charge qubit decoherence rate is incorporated through $\gamma$.
Importantly, the susceptibility is proportional to the difference in population between states $i$ and $j$, i.e. $\Delta p_{i,j} = p_i - p_j = \langle \sigma_z^{i,j} \rangle$ where $\sigma^{i,j}_z$ is the Pauli matrix for the respective transition. 
For more details see also reference \cite{Burkard2016}.

In the experiments presented above, a finite source-drain bias alters the DQD state population from what can otherwise be assumed to be an equilibrium thermal occupation $\propto e^{-E_{ij}/k_\mathrm{B}T}$.
The resulting steady-state non-equilibrium population is represented in the relative strength of the signal peaks within the bias triangles \cite{mi_high_2017, Viennot2014}. 
For our system, the internal and external quality factors of the resonator are $Q_\mathrm{int} \approx 7100$ and $Q_\mathrm{ext} \approx 2000$.
We estimate $g_0 / 2\pi \approx \SI{70}{\mega\hertz}$ through input-output theory, as well as a charge decoherence rate of $\gamma/2\pi \gtrsim \SI{300}{\mega\hertz}$.

\bibliography{mybibliography}

\end{document}